\documentclass[review]{elsarticle}

\usepackage{lineno,hyperref}
\usepackage{amssymb}
\usepackage{graphicx}
\usepackage{color}

\journal{Physica E}

\begin{document}
\begin{frontmatter}
\author{G. S. Diniz}
\address{Instituto de F\'isica, Universidade Federal  de Uberl\^andia, Uberl\^andia, MG 38400-902, Brazil}
\address{Curso de F\'isica, Universidade Federal de Goi\'as, 75801-615, Jata\'i, GO, Brazil}
\ead{ginetom@gmail.com}
\author{E. Vernek}
\author{F. M. Souza}
\address{Instituto de F\'isica, Universidade Federal  de Uberl\^andia, Uberl\^andia, MG 38400-902, Brazil}

\title{Graphene-based spin switch device via modulated Rashba Field and Strain}


\begin{abstract}
We investigate the spin-resolved transport in a two-terminal zigzag
graphene nanoribbon device with two independent gate induced Rashba spin-orbit
coupling regions and in the presence of strain. By employing a
recursive Green's function technique to the tight-binding model for the
graphene nanoribbon, we calculate the spin-resolved conductance of the
system. We show that by switching the sign of one of the gates it is
possible to select which spin component will be transmitted.
Moreover, our results show that an uniaxial strain applied to the
nanoribbon plays a significant role in the transport,
providing and additional manner to control the spin-polarized conductance.
This makes the present system a potential candidate for future
implementations of spin-based mechanical strain sensors.
\end{abstract}


\end{frontmatter}

\section{Introduction}
Fine control of electron spin degrees of freedom in nanostructures is
crucial for the development of future spin-based
electronics.\cite{RevModPhys.76.323}
With the advances in the growth and manipulation techniques at nanoscale, a
variety of materials have been suggested for future applications in spintronics
devices.\cite{Yongbing} In  particular, because of its
exceptional transport properties,\cite{Katsnelson} carbon-based materials
like graphene\cite{Novoselov22102004} have been considered as a
promising platform for spintronic devices,\cite{nnano2014214} attracting a
great deal of attention.\cite{Geim19062009,C5CP01637A,srep06464}

Pursuing the idea of designing spin-based devices, various schemes for
spin transistors have been proposed in the literature. The archetypal of
spintronic device is based on the  Datta-Das spin transistor.\cite{Datta-Das}
In graphene, two different spin-orbit couplings
(SOC) contributions  are present; the
\emph{intrinsic} and the \emph{extrinsic} SOC. The \emph{intrinsic}
SOC is known to be responsible for the
quantum spin Hall (QSH) phase in graphene: a time-reversal symmetry invariant
propagating gapless edge states. \cite{Kane,PhysRevLett.95.226801} If one of the spin
transport channels at the edges is suppressed, for instance by electron-electron
interaction \cite{JPSJ.65.1920,PhysRevB.84.115406}, or exchange field interaction,\cite{PhysRevB.85.115439} a
spin polarized edge-state is achieved and a phase transition from QSH to a quantum
anomalous Hall (QAH) phase is possible.\cite{PhysRevB.82.161414,PhysRevB.83.155447}

Because of the low atomic number of carbon atoms, \emph{intrinsic} SOC
is expected to be weak in pristine carbon nanostructures. However, there
are several studies demonstrating ways to enhance the intrinsic SOC, e.g.
by the proximity effect to transition metal dichalcogenides
(TDMCs),\cite{NatCom} or by adsorbed atoms on the graphene
surface.\cite{NatCom2,NatPhys} The \emph{extrinsic} SOC can be induced by
an external electric field, provided by underneath
gates.\cite{PhysRevLett.100.107602,Huertas,Zarea}

Spatially modulated Rashba fields have already been proposed in III-V semiconductor
quantum wires \cite{PhysRevB.69.035302,PhysRevB.80.041308,PhysRevB.84.075466,0953-8984-19-44-446209} and quantum rings. \cite{apl10.1063} In
Ref. \cite{PhysRevB.69.035302}, it was demonstrated
that depending on the electron doping, the spin polarized transmission can
be very sensitive to the width of both Rashba SOC and non-Rashba SOC
segments,\cite{PhysRevB.69.035302}, with appearance of spin-dependent conductance gaps.
In the same direction, the work presented in Ref. \cite{PhysRevB.80.041308,PhysRevB.84.075466}, a
metal to insulator state transition was observed in the case which the wave number of
the modulation commensurate with the Fermi wavelength of the injected electrons in the
quantum wire, demonstrating to be robust even in the presence of electron-electron interaction.
In Ref. \cite{apl10.1063} a finite chain of quantum
circular rings was used to investigate the electronic transport in the presence
of modulated Rashba SOC. They demonstrated that periodic modulations of
Rashba SOC were able to widen up transport gaps and produce an \emph{exotic}
nearly square-wave conductance \cite{apl10.1063}.

As compared to the III-V nanostructures, graphene-based devices have the advantage of
displaying several intriguing phenomena \cite{Katsunori} such as a versatile band
structure with localized edge states,\cite{PhysRevB.54.17954} zero-conductance Fano
resonances,\cite{PhysRevLett.84.3390} quantum spin Hall
effect,\cite{Kane,PhysRevB.80.115420,PhysRevB.92.075426} exceptional mechanical
properties \cite{Lee385} and spin-valley filtering.\cite{PhysRevLett.113.046601} All
these properties are handy for the control of their transport properties. Along
graphene-based structures, several proposals for using graphene as spin and valley filters
exploring SOC effects in a single barrier
have emerged in the past few years \cite{PhysRevLett.113.046601,PhysRevLett.112.136602,PhysRevB.91.165407}.

In this work we consider a simple device composed of a ZGNR deposited on top
of two spatially separated gates. These gates are used to induce
spatially modulated Rashba SOC.\cite{Zarea} By  using a Green's function
method, we calculate the differential conductance and show that by
appropriately tuning the Rashba SOC with the gate voltages, it
is possible to control the spin-polarized conductance of the system.
Moreover, we have studied the effect of an uniaxial strain along different
directions of the ribbon. We observe a strong dependence of the
spin-polarized conductance with strain, \cite{Diniz-JAP} by an appropriate
combination of gate voltages. The strain can modify the transport of a
selected spin component and can be used to design spintronic devices, e.g. spin-based
mechanical sensors.

\section{Theoretical Model}
\begin{figure}[!h]
\centering
\includegraphics[clip,width=3.1in]{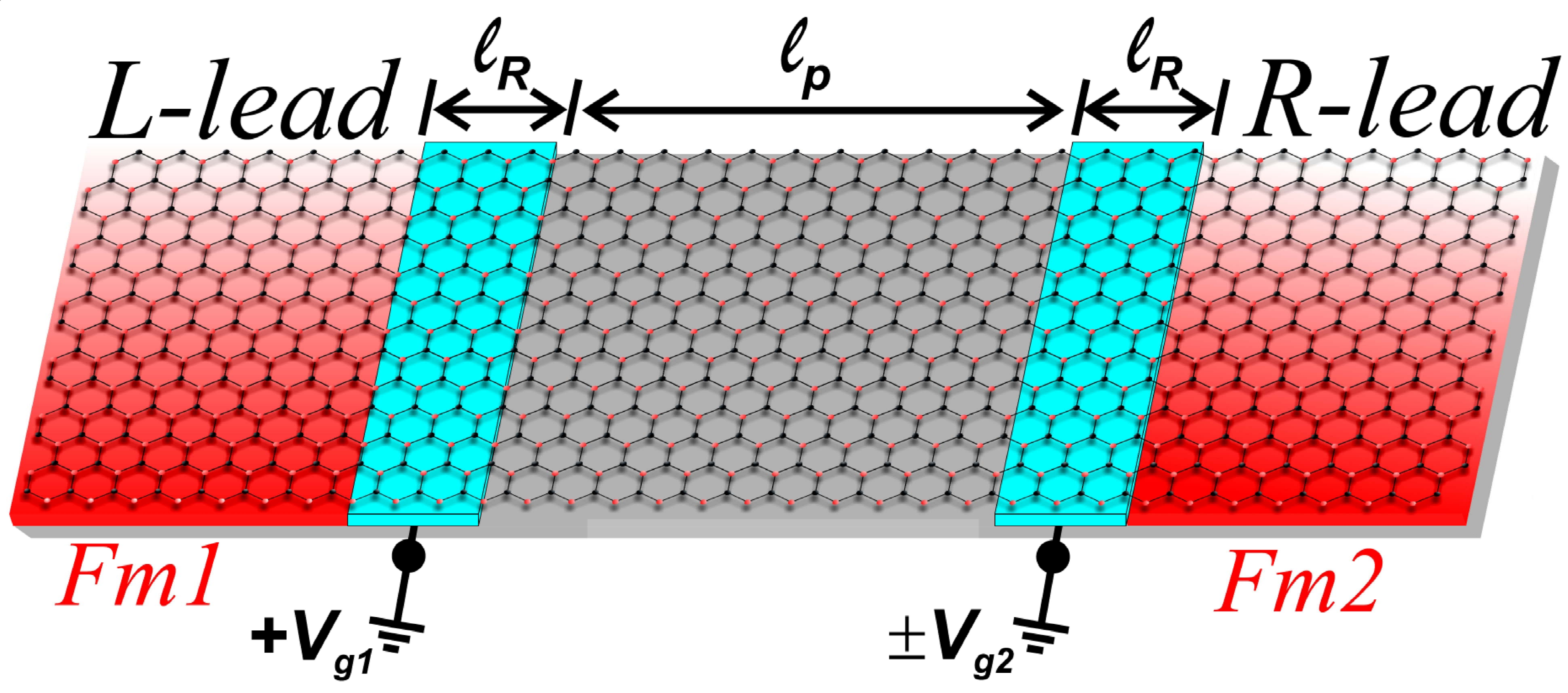}
\caption{Schematic representation of the nanoribbon device. A single layer
ZGNR is deposited on an appropriated substrate; local gates $V_{g1}$ and
$V_{g2}$ underneath the region $l_{R}$ (cyan) of the ZGNR  control the
Rashba SOC. The region $l_{p}$ (gray) between the gates is pristine ZGNR
width. Proximity effect to exchange coupled ferromagnetic insulators Fm1 and
Fm2 \cite{Swartz,PhysRevLett.112.116404,PhysRevLett.114.016603} or electric
fields \cite{Son,1.2821112} can be used to polarize the electrons in the ZGNR lead.}
\label{fig1}
\end{figure}

For concreteness, we consider a device
composed by a ZGNR deposited on the top of a substrate with underneath gates
that induce a Rashba SOC with tunable parameters $V_{g1}$ and $V_{g2}$
(see illustration  in Fig.~\ref{fig1}). The device is attached to pristine
semi-infinite ZGNR leads of identical chirality at both ends. To inject
(collect) spin-polarized electrons into (from) the device, an induced ferromagnetic ZGNR
lead can be used for this purpose. \cite{Swartz,PhysRevLett.112.116404,PhysRevLett.114.016603,0034-4885-73-5-056501,Son,1.2821112}

%
The modulated Rashba SOC system generated by the gates underneath the ZGNR (see
Fig. \ref{fig1}) is modeled using a $\pi$-orbital orthogonal tight-binding
Hamiltonian,
\begin{eqnarray}
\label{H0}
H\!=\! \sum_{\left< i,j
\right>\atop\sigma\sigma^{\prime}}\left[t_{ij}\delta_{\sigma\sigma^{\prime}} +
i \lambda_R\hat{z}\cdot(\vec{s}\times\vec{d}_{ij})
\right]c_{i\sigma}^{\dagger}c_{j\sigma^{\prime}} + H.c.,
\end{eqnarray}
where $c_{i\sigma}^{\dagger}$ ($c_{i\sigma}$) is the $\pi$-orbital creation
(annihilation) operator for an electron in the  $i$-th  site with spin
$\sigma$, $\vec{d}_{ij}$ is a lattice vector pointing from  $j$-th to the
$i$-th site of the ZGNR, $\vec{s}$ is a vector whose components are the
Pauli matrices and $\hat{z}$ is a unit vector perpendicular to the ZGNR
plane. $t_{ij}$=$t_{0}$ is the nearest neighbors hopping amplitude on the
honeycomb lattice. Finally, $\lambda_{R}$ represents the Rashba SOC
strength,\cite{Zarea} that is induced by the strong
electric fields generated  only within the region right above the gates
with width $l_{R}$, otherwise we set $\lambda_{R}$ to zero.

\emph{Strain effects} - In addition to the Rashba field we are also interested
in strain effects. The application of an external uniaxial stress to the ZGNR, or
the deposition of ZGNR on top of a substrate may induce an uniaxial strain.
\cite{NanoBao,apl3463460,nn800031m} To achieve a controllable uniaxial stress the graphene nanoribbons can be deposited
on flexible substrates, which can be perfectly stretched along specific
directions.\cite{Huang05052009,nn800459e,nl101533x,nl102123c} To simulate this uniaxial strain in our
system, we will consider a strain-dependent hopping parameter in our tight-binding
model.\cite{PhysRevB.80.045401} This simple model is capable of capturing the main
consequences of uniaxial strain on the band structure of graphene and
ZGNR.\cite{PhysRevB.81.035411,PhysRevB.88.085430,Lee18072008} Here the
strain modified distances between carbon atoms are described by
$\vec{d}^s_i=(I+\epsilon)\vec{d}_i$, with $\vec{d}_{i}$ (i= 1, 2, 3) the
unstrained vectors for nearest-neighbors, $I$ is the identity matrix and
$\epsilon$ is the strain tensor defined as \cite{PhysRevB.80.045401}
\begin{eqnarray}
\epsilon=
\varepsilon\left(\begin{array}{cc}
\cos^{2}\theta -\nu\sin^{2}\theta & (1+\nu)\cos\theta\sin\theta\\
(1+\nu)\cos\theta\sin\theta & \sin^{2}\theta -\nu\cos^{2}\theta\\
\end{array}\right).
\end{eqnarray}
Here, $\nu$ ($=0.165$) is the Poisson's ratio with the value known for
graphite,\cite{PhysRevB.80.045401} $\theta$ is the direction of strain
and $\varepsilon$ is the strain modulus. The hopping matrix element is
affected by the strain as $t_{ij}=t_{0} e^{-3.37(d^s_i/a_0-1)}$, in which
$t_{0}=2.7$eV is the unstrained hopping parameter\cite{PhysRevB.80.045401}
and $a_0$ (set as the unity) is the C-C distance. The $\theta$= 0
direction is parallel to the zigzag chain, and $\theta$= $\pi/2$ is
along armchair direction.

It is known that strain can induce band gap in armchair
GNR,\cite{Nanoresearch,Diniz-JAP} although no band gap is observed in
ZGNR.\cite{Diniz-JAP}
To illustrate the uniaxial strain effects on the conductance profiles of ZGNR
with modulated Rashba field, we assume that the entire system, composed by the
leads and the central conductor  are under the influence of stress, so that we
avoid any lattice mismatch at the interface.\cite{PhysRevB.88.195416}

\emph{Conductance--} To calculate the spin-resolved conductance, we use a
surface Green's function approach in real space.\cite{Nardelli} For this
purpose, we divide the two terminal ZGNR device into three well defined regions:
the left lead, the central conductor and the right lead. The central
conductor corresponds to the region with gates. The Rashba SOC
takes place only in the two regions on top of the gates. The Green's function
of the central conductor $\mathcal G_{C}$ (the spin index is omitted) is then
\begin{equation}
\mathcal G_{C}^{r/a}(E)
=\left(\omega_{\pm}-H_{C}-\Sigma_{L}-\Sigma_{R}\right)^{-1},
\end{equation}
where $a/r$ represents the advanced/retarded Green's function (with energy
$\omega_{\pm}=E\pm i\eta$, respectively, $\eta\rightarrow 0$), and $E$ is the
energy of the injected electron (the Fermi energy). $H_{C}$ denotes the
Hamiltonian in the central conductor and $\Sigma_{\mu=L, R}$ are the
self-energies for the connected left/right leads, $\Sigma_{\mu}=H_{\mu
C}^{\dagger}g_{\mu}H_{\mu C}$, where $g_{\mu}$ is the local
Green's function at the end of the semi-infinite left and right
leads.\cite{Nardelli} The matrix element $H_{\mu
C}$ gives the coupling between leads and central conductor. The spin-dependent
conductance through the central conductor is then calculated by,
\begin{equation}
G_{\sigma\bar{\sigma}}=G_{0}Tr\left[\Gamma_{\sigma}^{L}
\mathcal{G}_{C,\sigma\bar{\sigma}}^{r}\Gamma_{\bar{\sigma}}^{R}\mathcal{G}
_{C,\bar\sigma\sigma}^{a}\right],
\end{equation}
where the trace runs all the lattice sites in the central conductor. Here
$G_{0}=e^2/h$ is the quantum of conductance and $\Gamma_{\sigma}^{\mu}$
are the coupling matrices for the leads, associated to the spin-diagonal
self-energies $\Gamma^{\mu}=i\left[\Sigma_{\mu}^{r}-\Sigma_{\mu}^{a}\right]$. \cite{Datta,Nardelli}

When Rashba SOC is turned on in the device, the conductance profile will have
two different spin-dependent components: the spin-conserving component
$G_{\sigma\sigma}$), with $\sigma=\uparrow$ or $\downarrow$, that
is  associated with the injection and detection
of electrons with the same spin and the spin-flip component
($G_{\sigma\bar{\sigma}}$), resulting from spin rotation by the Rashba
field. It is important to mention that if there is no polarized electrons
being injected or drained by a ferromagnetic leads, time-reversal symmetry
is preserved in the device, therefore
$G_{\uparrow\uparrow}=G_{\downarrow\downarrow}$ and
$G_{\uparrow\downarrow}=G_{\downarrow\uparrow}$, resulting in no polarized
net current. The unpolarized electron flux can also be resulting of a multichannel
lead, as reported in Ref. \cite{PhysRevLett.94.246601} in two-terminal device.
For ferromagnetic leads, two possible configurations
can be used: (i) parallel alignment of the leads magnetization and (ii)
anti-parallel alignment. In the former case the transport is
carried on by electrons with the same spin in the source and drain, while
in the latter the transport is dominated by electrons with opposite spins
in the source and the drain.

\section{Numerical Results and Discussions}
%
Throughout this work we will assume a device composed of a 26-ZGNR (with width $(3/2N_{Z}-1)$,
$N_{Z}=26$) with a specific width of  $5.4{\rm nm}$ and length of
$21.9{\rm nm}$. For wider ribbons, there are more conducting channels available at
moderated energy of injected electron, henceforth, the device will increase
its complexity due to possible inter-channel scattering. Although, close to the Fermi
level there will be the same amount of conducting channels. We have also checked the
length dependence for a fixed region of Rashba field $\lambda_R$, and increasing
$l_P$, and it is indeed \emph{relevant} to the device prototype, but only
for higher energy doping (beyond 0.15$t_0$). Therefore, for electrons injected with
energies close to the Fermi level, the separation between the underneath gates is
\emph{irrelevant}, as the injected electron can not feel such fields in a
longer length scale beyond $\lambda_R$ region, which is a characteristic that
might be relevant in the experimental setup. Notice that although we choose a specific
width, the results presented show a general behavior of the ZGNR devices prototypes at low
energy regime, which is interesting for electronic transport. For \emph{metallic} armchair graphene
nanoribbons, we were able to obtain similar results in the absence of uniaxial strain, as depending on the
strain direction there is an induced transport gap close to the Fermi level, \cite{Diniz-JAP} which is not
our proposed effect: spin-selective filtering at low energy. It is important to mention that our formalism
is not able to capture possible valley filtering, as all the analysis is over the energy of injected electrons
that has contributions of both valleys (with no separation).

\begin{figure}[!h]
\centering
\includegraphics[scale=0.85]{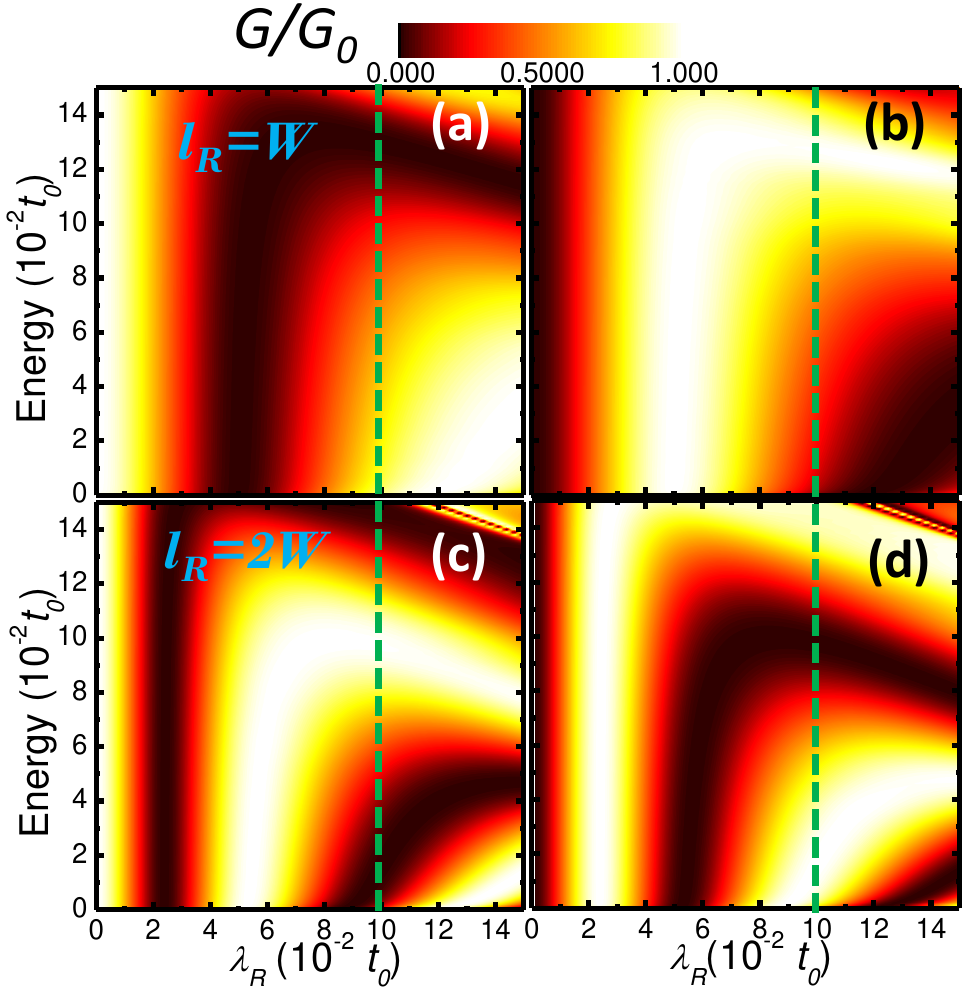}
\caption{Contour plots (Rashba parameter vs energy of injected electron) of the spin-dependent
conductance for different set of configurations. (a) Spin-conserving conductance $G_{\sigma\sigma}$
with width $l_{R}$=$W$, (b) spin-flip conductance $G_{\sigma\bar{\sigma}}$ with width
$l_{R}$=$W$. Panels (c) and (d) are for $G_{\sigma\sigma}$ and $G_{\sigma\bar{\sigma}}$, but
with different width $l_{R}$=$2W$. In all panels $V_{g1}=V_{g2}>0$ is assumed and we set the width of the SO region $W=2.46{\rm nm}$.}
\label{fig2}
\end{figure}

The contour plots of the spin-resolved conductance $G_{\sigma\bar{\sigma}}$ (Rashba
parameter vs energy of injected electron) is displayed in Fig. \ref{fig2}: (a) and (b)
spin conserving and spin-flip conductance with two gated regions with width $l_{R}=W$=10
unit cells long, respectively; (c) and (d) are similar but with twice the width $l_{R}=2W$=20
unit cells long. Hereafter, we set the width of the SO region $W=2.46{\rm nm}$ (corresponding
to 10 unit cells of the ZGNR). From the figures one can notice that while widening the $l_R$ width it causes
the reduction of the period in the oscillation pattern of the conductance profiles for fixed $E$, a
behavior attributed to the additional unit cells along the device with Rashba SOC field, which is
responsible for the electron spin precession along the device. It is also clear the opposite behavior
for the spin-conserving and spin-flip conductance; an enhancement in one of the components (brighter
region) reflects in the reduction of the other (dark regions). Another remarkable phenomenon is the
oscillatory dependence of the spin components of $G_{\sigma\bar{\sigma}}$ on the value of $\lambda_R$
for fixed $E$, which can be observed in all panels of Fig. \ref{fig2}. This oscillatory behavior is
reminiscent of the spin field effect transistor (FET) and has a similar origin,\cite{Datta-Das} as the
spin {\em precesses} as it propagates in the presence of the Rashba field, acquiring a net phase that is
proportional to $\lambda_{R}$ and the total length of the central conductor.

\begin{figure}[t!]
\centering
\includegraphics[clip,width=3.2in]{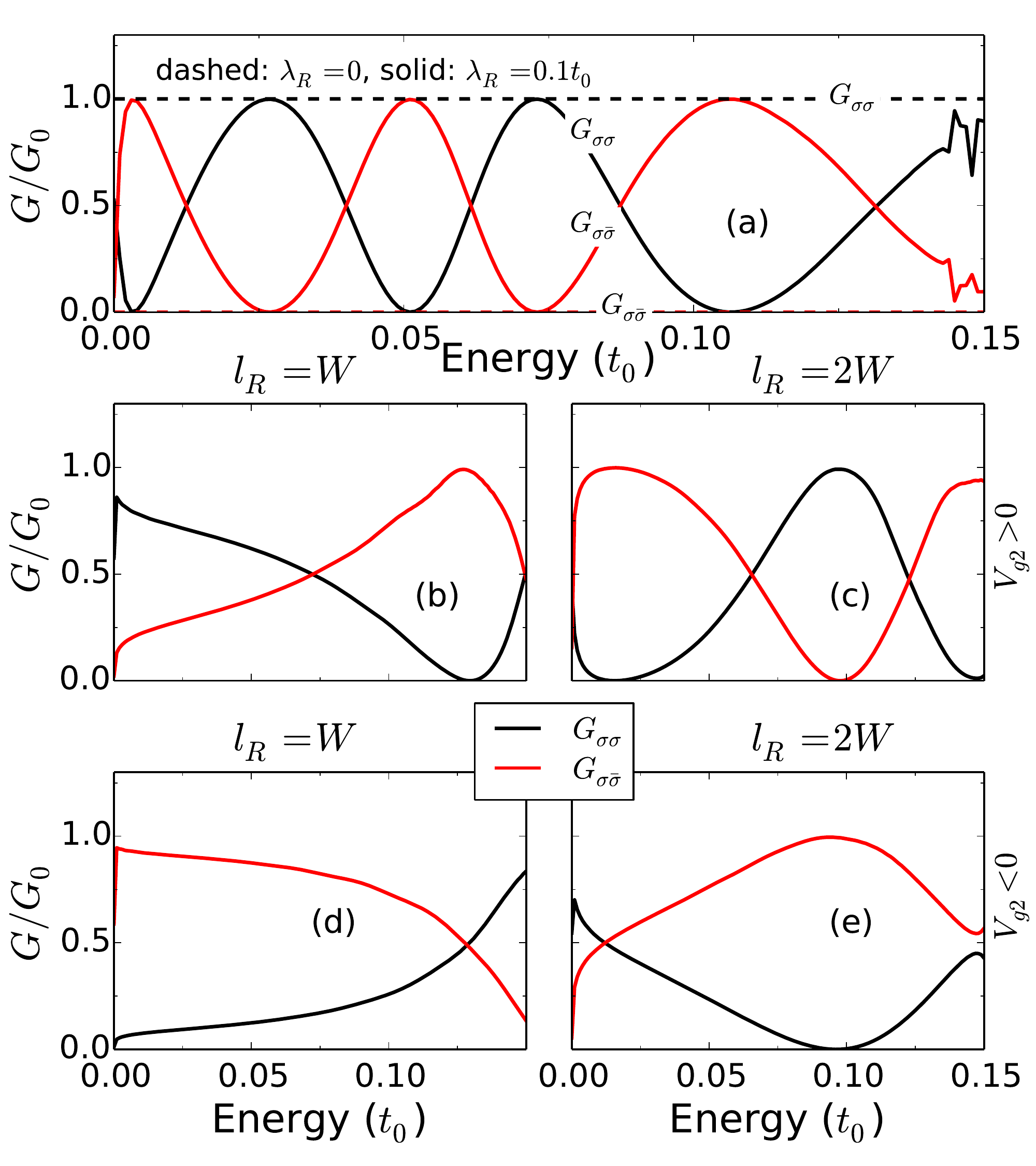}
\caption{(Color online) Spin dependent conductances
$G_{\sigma\sigma}$ (black) and $G_{\sigma\bar\sigma}$ (red) vs energy. (a)
$\lambda_{R}=0$ (dashed) $\lambda_{R}=0.1t_{0}$ (solid) for
$V_{g1}=V_{g2}$ applied to the entire region underneath the device. (b)
and (c)  for width $l_R=W$ and $l_R=2W$, respectively; (d) and (e) similar
(b) and (c) but for $V_{g2}<0$, respectively. For all panels (b)-(e),
$\lambda_{R}$=0.10$t_{0}$.}
\label{fig3}
\end{figure}

In what follows we set the Rashba parameter
$\lambda_R=0.1 t_0$ and $V_{g1}=V_0$ (vertical dashed line in Fig. \ref{fig2}), that produces a
z-dependent potential $V(z)$, leading to a Rashba SOC
$\lambda_R=-(e/2m^2v_{f})(dV/dz)$, where $dV/dz={\tt E}$ is the electric
field perpendicular to the ZGNR plane.\cite{Zarea}
We start by  showing in Fig.~\ref{fig3}(a) the conductance vs
energy for the situation in which the gate voltage in applied to the
entire region underneath the device. For $\lambda_R=0$ (dashed lines) we
see that $G_{\sigma\sigma}=G_0$  for the entire range of energy shown
while $G_{\sigma\bar\sigma}=0$. When $\lambda=0.1t_0$, however both
conductances oscillate with opposite phase (note that the maximum of
$G_{\sigma\sigma}$ corresponds to the minimum of $G_{\sigma\bar\sigma}=0$
and vice versa). These oscillations result from the spin rotation produced
by the SOC. The period of oscillation is due to
competing effects. As the energy increases the SOC becomes more
pronounced, however the electrons travel faster across the device, having less time to precess.
This turns into slower oscillations in the conductance for increasing
energy. \cite{Datta-Das} Besides, for larger energies more conducting
channels contributes to the conductance.

The system  is more tunable in the case of two independently gated
regions. In Fig.~\ref{fig3}(b)  and \ref{fig3}(c) we show the spin
dependent conductances  $G_{\sigma\sigma}$ and
$G_{\sigma\bar\sigma}$ vs
energy for $l_R=W$ and $l_R=2W$, respectively and for $V_{g2}=V_0$. When comparing these results with those of
Fig.~\ref{fig3}(a) we see that for $l_R=W$ the conductances oscillate much
slower with the energy because the influence of the SOC is smaller
(remember that while traveling across the central region $l_p$ the
electrons preserve their spin because there is no SOC in this region).
Note that when $l_R$ increase for $l_R=2W$ [as in Fig.~\ref{fig3}(c)] the
profile of the curves approaches that of the Fig.~\ref{fig3}(a).

\begin{figure}[b!]
\centering
\includegraphics[clip,width=3.2in]{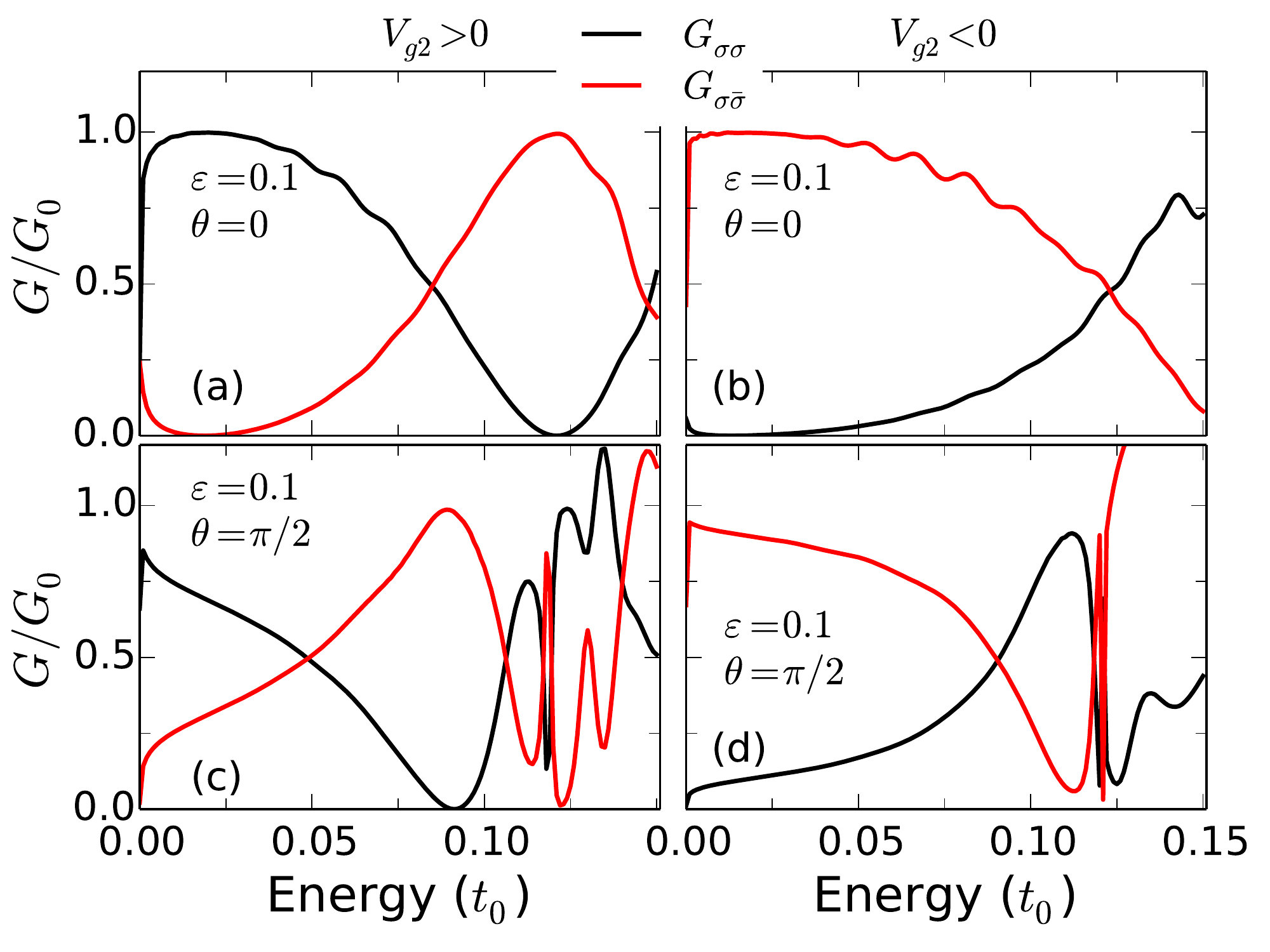}
\caption{(Color online) $G_{\sigma\sigma}$ (black) and
$G_{\sigma\bar\sigma}$ (red)  vs $E$ for strained ZGNR with
$\varepsilon$=0.1 and $\theta$=0 for $V_{g2}>0$ and
$V_{g2}<0$, respectively. (c) and (d) are the results for the same
values of parameter as in the panels (a) and (b) except that
$\theta=\pi/2$. For all cases we fixed
 $l_{R}$=$W$ and $\lambda_{R}$=0.1$t_{0}$.}
\label{fig4}
\end{figure}

Now, in Fig.~\ref{fig3}(d) and \ref{fig3}(e) we switch the sign
of $V_{g2}$ such that $V_{g2}=-V_{g1}=-V_0$. In this case, by comparing
these results with those of Figs.~\ref{fig3}(b) and
\ref{fig3}(c) we note a completely different profile. Note, for
example, that the behavior of  $G_{\sigma \sigma}$ and
$G_{\sigma\bar\sigma}$ are inverted. This is best seen for $l_R=2W$
[Figs.~\ref{fig3}(c) and \ref{fig3}(e)]. Observe
that for $E\approx 0.1t_0$, while $G_{\sigma \sigma}\approx G_0$  and
$G_{\sigma\bar\sigma}\approx 0$ for $V_{g2}=V_0$, the opposite occurs
for $V_{g_2}=-V_0$. Naively, one could think that the spins of
the electrons rotated by $V_{g1}$ in the first region would
be rotated back to their original orientation when going through the
region over $V_{g2}$. This switch between $G_{\sigma \sigma}$
and $G_{\sigma\bar\sigma}$ as $V_{g2}$ changes sign can be useful
to control the current intensity when both leads are
ferromagnetic.
For instance, when the magnetization of the leads are parallel aligned, the
transport will be dominated by $G_{\sigma \sigma}$  so the current will be
higher for $V_{g2}>0$ than for $V_{g2}<0$. For antiparallel
alignment of the leads magnetization, we expect the opposite, i.e., higher current for $V_{g2}<0$ than
for $V_{g2}>0$, since in this configuration the transport will be dominated
by $G_{\sigma\bar\sigma}$.

\begin{figure}[!h]
\centering
\includegraphics[clip,width=3.4in]{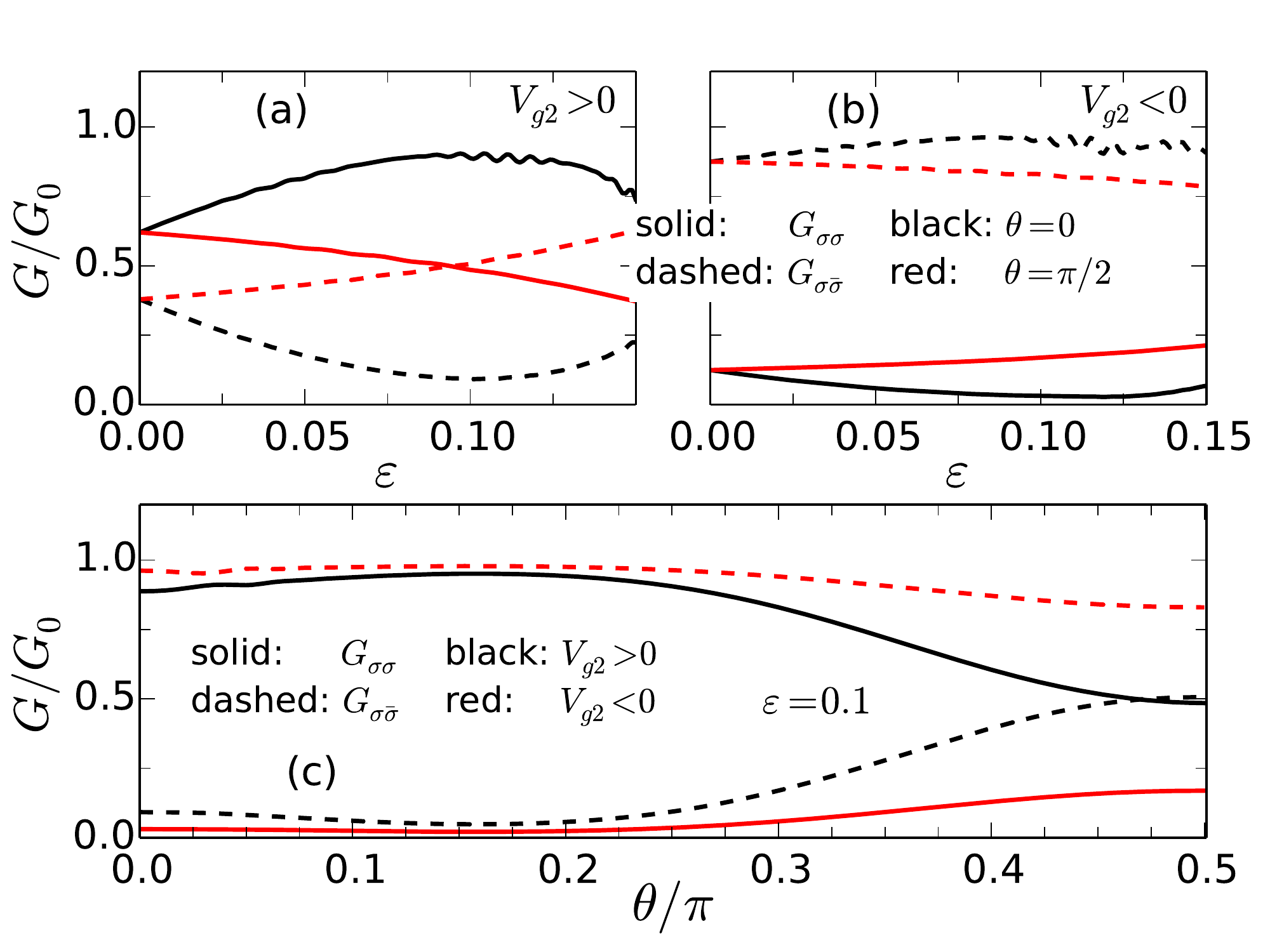}
\caption{(Color online) (a)-(b) Conductance as function of $\varepsilon$ for fixed applied
direction of strain $\theta$=0 and $\pi/2$, for $V_{g2}>0$ and $V_{g2}<0$,
respectively. (c) conductance vs  $\theta$ for fixed $\varepsilon $=0.1
with two possible configurations of the gates. For all cases we fixed
 $l_{R}$=$W$, $\lambda_{R}$=0.1$t_{0}$ and $E=$0.05$t_{0}$.}
\label{fig5}
\end{figure}

Another appealing behavior of our device is revealed when it is under
strain. To appreciate this, in Fig.~\ref{fig4} we show the effect of the
uniaxial strain effects on the spin-polarized conductance. The strain
parameter is taken as  $\varepsilon =0.1$ in two different directions: (i)
$\theta=0$ (along the ribbon) and (ii) $\theta=\pi /2$ (transversal to the
ribbon). Here only $l_R=W$ is considered. For
$\theta=0$ [Fig.~\ref{fig4}(a)] the conductances $G_{\sigma\sigma}$ and
$G_{\sigma\bar\sigma}$ preserve their main features seen in the
unstrained case, for both positive and negative $V_{g_2}$. For a strain
along $\theta=\pi /2$ [Fig.~\ref{fig4}(c)] the variation of
 $G_{\sigma\sigma}$ and $G_{\sigma\bar\sigma}$ with energy is faster as
compared with the unstrained case. Because for $\theta=\pi/2$ the
conductance is more sensitive to the energy we note also that additional
structures in the conductance are seen for higher energies that were note
seen in the case of $\theta=0$. Physically, this is because
a positive $\varepsilon$ along $\theta=\pi/2$ increases the width of the
ribbon (for a fixed number of carbon atoms), reducing the
transversal confinement.

In Fig.~\ref{fig5}(a) and \ref{fig5}(b) we also fixed the energy  at
$E=0.05t_0$ and plot the conductances $G_{\sigma\sigma}$ (solid)
 and $G_{\sigma\bar\sigma}$ (dashed) vs strain $\varepsilon$,
respectively, for  $\theta = 0$ (black) and $\theta = \pi/2$ (red). For
$V_{g2}=V_0$ and $\theta = 0$, as $\varepsilon$ increases the
spin-conserving conductance $G_{\sigma\sigma}$ increases while the
spin-flip component $G_{\sigma\bar\sigma}$ decreases. The opposite is seen
for $\theta=\pi /2$. For $V_{g_2}=-V_0$ [Fig. \ref{fig5}(b)], although the
conductances are less sensitive to $\varepsilon$, the picture is
fully reversed. In this situation $G_{\sigma\sigma}$
($G_{\sigma\bar\sigma}$) decreases (increases) as the strain enhances along
$\theta = 0$, while for $\theta=\pi /2$, $G_{\sigma\sigma}$
($G_{\bar\sigma}$) increases (decreases) with strain.

Finally, figure \ref{fig5}(c) shows how  $G_{\sigma\sigma}$ and
$G_{\sigma\bar\sigma}$ evolves with $\theta$ and fixed strain parameter
$\varepsilon=0.1$. For $V_{g_2} > 0$ (black lines) $G_{\sigma\sigma}$ remains
close to one, while $G_{\sigma\bar\sigma}$ is close to zero. As strain is
rotated
both $G_{\sigma\sigma}$ and $G_{\sigma\bar\sigma}$ approaches the same
value
around $0.5G_{0}$. For $V_{g_2}=-V_0$ the conductance $G_{\sigma\sigma}$
($G_{\sigma\bar\sigma}$) is only slightly suppressed (enhanced).
These
features could be exploited as a possible manner to control the spin-polarized
transport via strain, with potential application as a spintronic mechanical strain sensors.
%

\section{Conclusion}
Summarizing, we have studied the spin
transport through ZGNR device composed of two local gates with Rashba SOC
field generators. We demonstrate the feasibility of spin-charge current flow
manipulation by means of the spatially modulate Rashba field and strain.
By changing  the sign of the underneath gates, it is possible to control the
spin selective current intensity by considering spin-polarized leads.
Furthermore, we also show that strain plays an important role in the
spin polarized current control being an additional tuning parameter.
In the light of our numerical simulations, we expect our results to be of great
applicability in the GNR spin-based devices as well as for other 2D-related materials in
which similar setups can be performed.

\emph{\textbf{Acknowledgements}--}We acknowledge financial support received
from FAPEMIG, CAPES and CNPq.




\section*{References}
\bibliography{references} 
\end{document}